\documentclass[twoside,12pt,journal]{IEEEtran}
\normalsize

\usepackage[pdftex]{graphicx}
\usepackage[cmex10]{amsmath}
\usepackage{amssymb}
\usepackage[caption=false,font=footnotesize]{subfig}
\usepackage{booktabs}

\usepackage{url}
\usepackage{mathrsfs}
\usepackage{glossaries}

\setglossarystyle{treenoname}
\usepackage{mathtools}
\usepackage{xparse}
\usepackage{cite}
\usepackage{color}

\newcommand{\round}[1]{\ensuremath{\left\lfloor#1\right\rceil}}

\def\hSSF	{\mathit{h}_\mathrm{T}}
\def\hSSFv	{\mathbf{h}_\mathrm{T}}
\def\w	{\mathbf{w}}
\def\HSSF	{\mathit{H}_\mathrm{T}}
\def\D		{\mathit{D}_\mathrm{T}}
\def\V{\mathbf{V}}
\def\dT	{\mathbf{d}_\mathrm{T}}
\def\st{\mathop{\mathrm{subject~to}}}
\def\max{\mathop{\mathrm{max}}}

\def\minimize{\mathop{\mathrm{minimize}}}

\def\argmax{\mathop{\mathrm{argmax}}}

\DeclareMathOperator{\EX}{\mathrm{E}} 
\DeclarePairedDelimiter\abs{\lvert}{\rvert}
\newcommand{\norm}[1]{\left\lVert#1\right\rVert}

\newcommand{\ju}{{\mathrm{j}\mkern1mu}}
\def\e {\mathrm{e}}
\def\phimax{{\varphi_{\max}}}
\def\Sphi{{K_{\varphi}}}
\def\Iphi{{{\cal I}_{\varphi}}}
\def\Rrrc{{R_{\text{RRC}}}}
\def\Rssf{{R_{\text{SSF}}}}
\def\Trrc{{T_{\text{RRC}}}}

\NewDocumentCommand{\INTERVALINNARDS}{ m m }{
	#1 {,} #2
}
\NewDocumentCommand{\interval}{ s m >{\SplitArgument{1}{,}}m m o }{
	\IfBooleanTF{#1}{
		\left#2 \INTERVALINNARDS #3 \right#4
	}{
	\IfValueTF{#5}{
		#5{#2} \INTERVALINNARDS #3 #5{#4}
	}{
	#2 \INTERVALINNARDS #3 #4
}
}
}
\newcommand{\cmmnt}[1]{\ignorespaces}

\begin{document}

\title{Efficient Spectrum Utilization via Pulse Shape Design for Fixed Transmission Networks}

        
\author{Elena-Iulia~Dobre$^*$, Ayman~Mostafa$^*$, 
        Lutz~Lampe$^*$,
        Hoda~ShahMohammadian$^\dagger$,
        Ming~Jian$^\dagger$

\thanks{$^*$Department
of Electrical and Computer Engineering, The University of British Columbia, Vancouver, Canada}%
\thanks{$^\dagger$Huawei Technologies, Ottawa, Canada}%
\thanks{Resubmission. Original manuscript submitted to IEEE Transactions on Communications on 15 November 2019.}%
}

\maketitle
\thispagestyle{empty}
\addtocounter{page}{-1}

\begin{abstract}
Microwave backhaul links are characterized by high signal-to-noise ratios permitting spectrally-efficient transmission. The used signal constellation sizes and achievable data rates are typically limited by transceiver impairments, predominantly by phase noise from non-ideal carrier generation. In this paper, we propose a new method to improve the data rate over such microwave links. We make use of the fact that adjacent frequency channels are inactive in many deployment scenarios. We argue that additional data can be transmitted in the skirts of the spectral mask imposed on the transmission signal by regulation. To accomplish this task, we present a shaped wideband single-carrier transmission using non-Nyquist pulse shapes. In particular, we design spectrum-skirt filling (SSF) pulse shaping filters that follow the spectral mask response, and perform detection using an accordingly increased sampling frequency at the receiver. We evaluate the achievable information rates of the SSF-based transmission considering practical dispersive channels and non-ideal transmitter and receiver processing. To compensate for phase noise impairments, we derive carrier phase tracking and estimation techniques, and utilize them in tandem with nonlinear precoding which mitigates the intersymbol interference introduced by the non-Nyquist SSF shaping filter. Quantitative performance evaluations show that the proposed system design achieves higher data rates in a dispersive microwave propagation environment with respect to the conventional transmission with Nyquist pulse shaping.    
\end{abstract}


\begin{IEEEkeywords}
Shaped wideband transmission, pulse shaping filter, Tomlinson-Harashima precoding, phase noise.
\end{IEEEkeywords}

%
\IEEEpeerreviewmaketitle

\section{Introduction}\label{Introduction}
%
%
%
%

Microwave transmission systems have widely been used for fixed broadband and mobile backhaul networks. For example, about 50\%-60\% of the base stations worldwide are connected through microwave links \cite{lombardi2015,Ericsson_MicrowaveTowards2020}. The data rate requirements for these links are continuously increasing to support the fast growing demands of radio access networks. Improving the data rate of fixed microwave links has a long history \cite{Kizer2013}. Modulation formats have evolved from \(128\text{-ary}\) quadrature amplitude modulation (QAM) \cite{itu2005} to \(1024\text{-QAM}\) \cite{itu2015}, and lately up to \(8192\text{-QAM}\) \cite{huawei2016}. Microwave systems use large constellation sizes as part of adaptive modulation schemes, which select the modulation depending on the fairly slowly varying signal-to-noise ratio (SNR) of the communication link. The throughput of microwave links can be further improved by the use of dual-polarization and spatial multiplexing, where the latter requires an appropriate antenna arrangement to achieve a multiplexing gain in line-of-sight conditions \cite{cvetkovski2016}. Another possibility is the use of additional frequency bands. While fixed microwave systems have commonly operated in frequency bands from 6~GHz to 42~GHz, more recently, millimeter wave bands, in particular the 57-66 GHz V-band and the 71-86 GHz E-band, have been used \cite{Ericsson_MicrowaveTowards2020,itu2015,huawei2016}.

\begin{figure}[t]
	\centering
	\includegraphics[width=1\columnwidth]{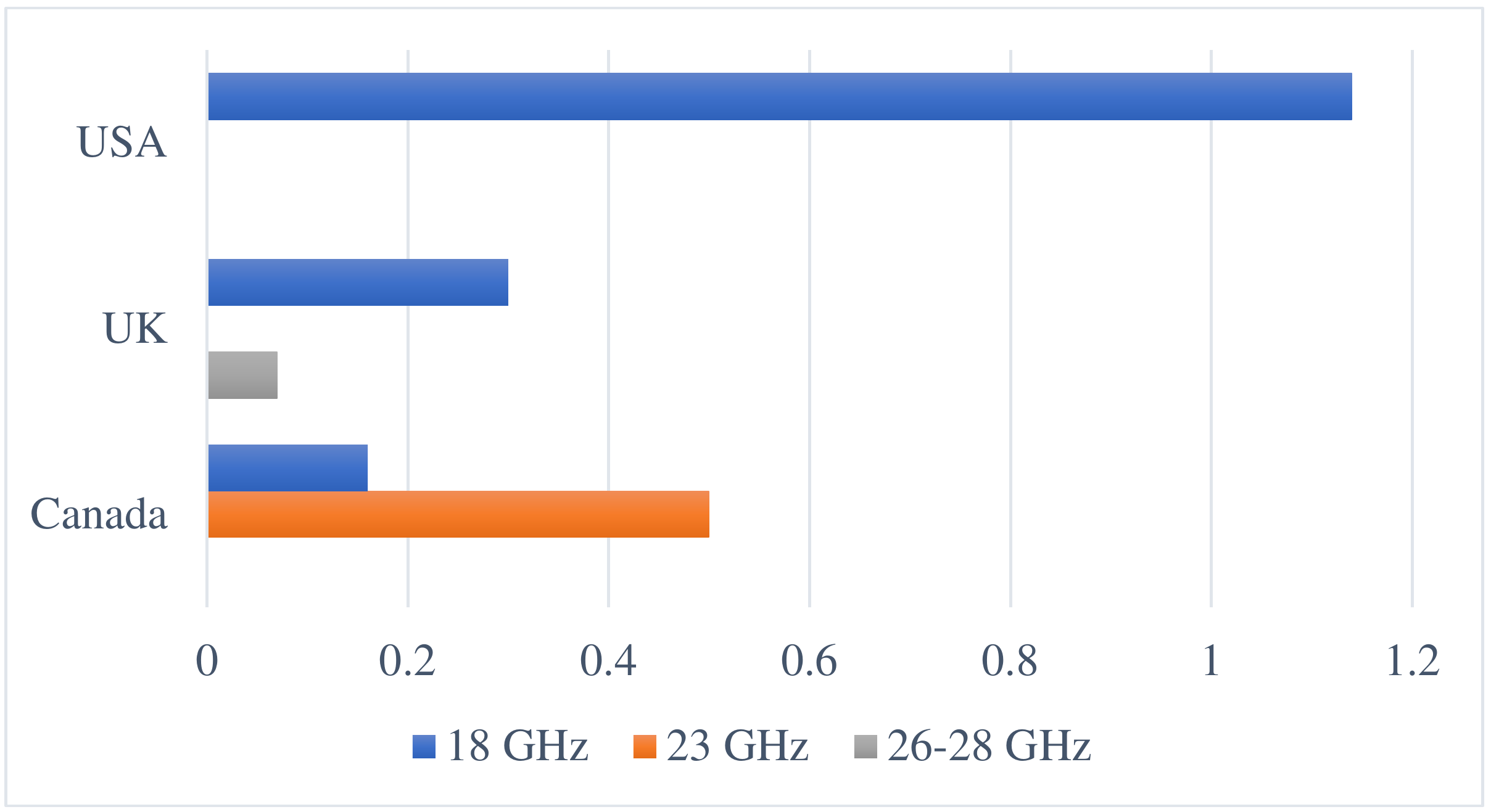}
	\caption{\textcolor{black}{Percentage of fixed point-to-point microwave frequency channels with active neighbours in the same transmission hop.}}
	\label{Channels_with_active_neighbours}
\end{figure}

In this paper, we propose a method to enhance the data rate of microwave transmission systems that is complementary to all of the above-mentioned approaches. The essence of our approach is to replace conventional waveforms, which utilize a Nyquist pulse in the main lobe of the spectral mask, by a pulse shape that fills the entire area of the mask, including the spectrum skirts.
The main motivation of the proposed spectrum-skirt filling (SSF) wideband transmission derives from the numerous instances in which the adjacent frequency channels are not utilized in the same transmission hop.
\textcolor{black}{To demonstrate the feasibility of transmission in the skirts, we computed the fraction of channels with active neighbours in the same hop, based on the latest frequency allocation tables for Canada \cite{SpectrumManagementSystemData}, the UK \cite{WirelessTelegraphyRegister}, and the USA \cite{UniversalLicensingSystem}. Figure~\ref{Channels_with_active_neighbours} shows the results in percent for several microwave bands for point-to-point fixed links. 
The very low neighbour-activity rates demonstrate the potential for using the spectrum skirts to achieve higher data rates.}
\textcolor{black}{Our second premise is the high SNR experienced by fixed microwave links for \(99\% - 99.9\%\) of the time of operation \cite{Huang_MultiGigabitMicrowave}.
This suggests data transmission with non-negligible rates is possible in spectrum skirts with a power spectral density (PSD) of 35-45~dB below the level at the carrier frequency.}
It follows that the regulated spectrum skirts could be used to transmit additional data without the need to lease another channel, and thus without additional costs for the setup of a microwave link, including costs for spectrum licenses.

A contrasting method of improving the data rate in a conventional root-raised cosine (RRC) system is increasing the spectral efficiency.
Although commercial systems use high-resolution analog-to-digital converters to process large constellations, phase noise (PN) sets a constraint on the QAM alphabet size due increased PN sensitivity for higher-order modulation \cite{Kreimer_EfficientIterativePhaseEstDecoding}. This can partially be addressed by effective PN compensation techniques, often devised jointly with equalization \cite{Stark_CombiningDFEandCPE}, or as iterative trellis-based PN compensation and decoding \cite{Peleg_IterativeDecoding, Colavolpe_IterativeDecodingPhaseNoise}. In this context, it is insightful to compare the possible relative rate improvement from the use of an increased constellation size with that from using spectrum skirts. Replacing \(4096\text{-QAM}\) by \(8192\text{-QAM}\) leads to an 8\% rate increase, assuming that a reliable detection is possible. Using instead the spectrum skirts provided by a spectral mask with 28~MHz channel spacing (details will be provided in Section~\ref{PulseShapingFilterDesign} and Figure~\ref{MagResp_SSF_RRC_Mask}), and considering a typical link SNR of 50~dB at the carrier frequency, an improvement of 50\% in data rate is possible according to the communication-theoretic limit for transmission with a given PSD \cite{Anderson_FTNSignaling}.

\textcolor{black}{We note that enhanced RRC pulse shape designs such as the ones considered in \cite{Muravchik_OptimizedSignaling, Assimonis_TwoParamNyquistPulses, FarhangBoroujeny_SquareRootNyquist} for improving robustness against timing jitter, in \cite{Baas_PulseShapingFreqSelCh} to match the pulse shaping filter to the channel characteristics, or in\cite{Eghbali_TwoStageNyquist} to minimize the multiplicative complexity of the pulse shaping filter implementation would not improve the rate of the RRC benchmark system in our microwave transmission link. In particular, we assume accurate timing synchronization, an adequate number of taps for pulse shaping and equalization filters, and a mild intersymbol interference (ISI) channel, for which the conventional RRC pulse provides the best performance, and thus is the appropriate benchmark to compare against.}
The proposed shaped transmission has similarities to Faster-than-Nyquist (FtN) signaling \cite{Anderson_FTNSignaling} through enabling faster rates at the cost of losing pulse orthogonality.
\textcolor{black}{However, unlike FtN, which starts from a Nyquist pulse and in essence uses the excess bandwidth, i.e., the difference between transmission rate and pulse shape bandwidth, to increase data rate, our SSF transmission shapes the signal PSD.
This, in turn, suggests much higher potential rate improvements than for FtN. For example, for the case of an RRC pulse with a roll-off factor $\beta$, the achievable rate gain for FtN would be limited to the factor of $(1+\beta)$. For SSF, the potential rate gain is determined by the achievable rate given the spectrum mask.}

\textcolor{black}{Several works propose the design of pulse shaping filters which conform to spectral mask constraints.
In \cite{Davidson_LMI_SpectralMask}, the objectives of the design are to minimize the stopband energy for a fixed filter length, to trade-off the ISI introduced by the filter and the channel distortion, or to minimize the bandwidth for a fixed admissible ISI.
Similarly, in \cite{Dedeoglu_FilterDesignDIRR}, the authors propose a design which minimizes the stopband energy of the filter under frequency response constraints.
In \cite{Davidson_OrthogonalPulseShapes_SP}, an orthogonal multirate filter which concentrates most of the pulse's energy in a minimum bandwidth for a given length, or in a minimum filter length for a fixed bandwidth, or which maximizes the spectral energy concentration within most of a given bandwidth and fixed length is presented.
Other avenues which make use of a spectral mask to find a pulse shaping filter, albeit for entirely different use cases, are proposed in \cite{Macaluso_BEMCompliantWaveforms} and \cite{Beaulieu_PulseShape_UWB}.
The work in \cite{Macaluso_BEMCompliantWaveforms} shapes orthogonal frequency-division multiplexing waveforms to be compliant to a block-edge mask to enable dynamic spectrum access while suppressing out-of-band radiation.
The filter design in \cite{Beaulieu_PulseShape_UWB} maximizes the signal-to-noise-and-interference ratio while meeting the mask constraints in a multiple access ultra-wideband communication system.}
\textcolor{black}{What differentiates our paper from these previous works is the pulse shape design with the express purpose of exploiting spectrum skirts for increasing data rates in the case of vacant adjacent channels. Furthermore, our focus is not on resource optimality, such as filter length or signal energy concentration, for which we could however adopt existing methods if desired.} 

The proposed SSF transmission introduces ISI. Hence, we apply several equalization methods that work adequately with PN compensation. For the latter, we focus on low-complexity carrier phase tracking via a digital phase-locked loop (DPLL), and PN estimation via the Bahl-Cocke-Jelinek-Raviv (BCJR) algorithm \cite{BCJR_OptimalDecoding}, building on the work in \cite{Stark_CombiningDFEandCPE, Hoda_AccurateBCJRbasedCPE}. 
\textcolor{black}{Our PN compensation schemes extend these methods to operate in tandem with a favourable pre-equalization technique.}
Against this background, our main contributions are summarized as follows. 
\begin{itemize}
\item We propose a new transmission waveform concept, which can be applied in the numerous microwave links whose frequency-adjacent channels are not occupied. The magnitude response of the designed pulse shaping filter fills the spectrum skirts of the regulated mask and supports a higher data rate due to the higher bandwidth of the transmitted waveforms.
\item We show that low-complexity equalization at the receiver side is not well suited to combat the ISI stemming from the SSF filter. Thus, using the fact that the ISI is known \textsl{a priori}, we apply Tomlinson-Harashima precoding (THP) \cite{Tomlinson_THP, Harashima_THP} at the transmitter. Furthermore, we adjust the PN compensation methods to operate on the extended signal constellation induced by THP and to yield reliable estimates of the PN at pilot symbols.
\item We utilize the achievable information rate (AIR) \cite{Arnold_InformationRate, Eriksson_AchievableRates}, symbol error rate (SER), \textcolor{black}{and bit error rate (BER) of the transmission with forward error correction (FEC)} to analyze the performance of our transceiver. \textcolor{black}{The performance results of our simulated transmission scenarios} show that the proposed transmission technique achieves significant data rate gains \textcolor{black}{from 23\% to 40\%} in comparison to the standard RRC-based transmission, depending on the applied PN compensation method.
\end{itemize}

The remainder of this paper is organized as follows. In Section~\ref{OverallSystemCharacterization}, we describe the proposed system architecture, followed by a summary of the pulse shaping filter design in Section~\ref{PulseShapingFilterDesign} and equalization and pre-equalization methods in Section~\ref{SimplifiedReceiverDSP}. In Sections~\ref{NonlinearPrecoding_DPLL_CPE} and~\ref{NonlinearPrecoding_BCJR_CPE}, we derive the details of the two proposed joint precoding and PN compensation techniques. Section~\ref{NumericalResultsDiscussion} presents a quantitative performance evaluation of the proposed system design and detection methods. Section~\ref{Conclusions} concludes this paper.

\textit{Notation:}
The absolute value, real part, imaginary part, complex conjugate, and expectation are denoted by \(\abs{\cdot}\), \(\Re\{\cdot\}\), \(\Im\{\cdot\}\), \({(\cdot})^*\) and \(\EX\{\cdot\}\), respectively. Rounding towards the nearest integer is denoted by \(\round{\cdot}\), \textcolor{black}{\(\odot\) and \(\left(\cdot\right)^{\abs{\cdot}}\) denote element-wise multiplication and element-wise absolute value, respectively}.

\section{Proposed System Design}\label{ProposedSystemDesign}

In this section, we describe the proposed system layout, and focus on the pulse shaping filter design, as well as the equalizer structure, which aims to mitigate the ISI due to filtering and propagation through a dispersive channel.

\subsection{Overall System Characterization}\label{OverallSystemCharacterization}

\begin{figure*}[!t]
	\centering
	\subfloat[Transmitter\label{TransmitterInitDiagram}]{
		\includegraphics[width=0.8\textwidth]{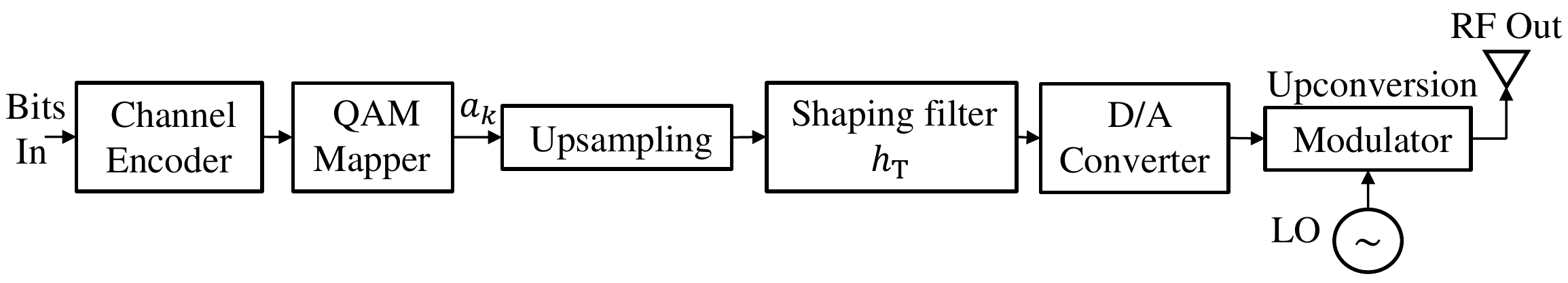}
	}
	\vfill
	\subfloat[Receiver\label{ReceiverInitDiagram}]{
		\includegraphics[width=0.8\textwidth]{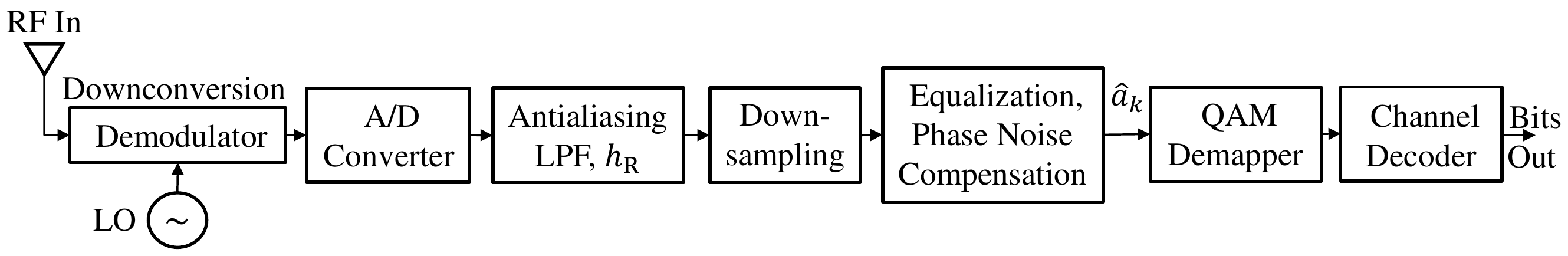}
	}
	\caption{\textcolor{black}{Proposed system design for an efficient spectrum utilization transmission.}}
	\label{TxAndRxInitDiagrams}
\end{figure*}

We propose a spectrally efficient communication system with the transmitter and receiver structures shown in Figure~\ref{TxAndRxInitDiagrams}. At the transmitter side, the input bits
\textcolor{black}{are FEC encoded and} mapped into $M$-ary QAM symbols followed by upsampling and pulse shaping with the SSF filter \(\hSSF\). 
The SSF filter constitutes a key component of the system design and will be discussed in detail in Section~\ref{PulseShapingFilterDesign}.
The baseband analog pulse is up-converted to a carrier frequency fit for microwave transmission. Due to hardware imperfections of the local oscillator (LO), the generated carrier is non-ideal and includes PN. This impairment causes random rotations of symbols, which leads to detection errors and, if uncompensated, affects the overall system performance.

The transmitted signal experiences distortion due to the microwave channel, for which we adopt the widely used \textcolor{black}{three-path} Rummler model, detailed in \cite{Rummler_MultipathFading}. The proposed receiver design has a direct-conversion architecture, where the LO generated carrier frequency is equal to the one of the desired signal. At receiver, the signal in the desired frequency band is down-converted to baseband. As with the transmitter LO, the carrier generation at the receiver LO is impaired by PN. Following digitization, \textcolor{black}{the filtered and downsampled} symbol sequence is processed by an interference and impairment mitigation block consisting of equalization and PN compensation, mapped back to bits \textcolor{black}{and FEC decoded}.

\textcolor{black}{The received symbol sequence of the equivalent discrete-time complex baseband model can be expressed as}
\begin{equation}\label{DiscreteBasebandSystemEq}
	y_{k} = \e^{\ju \varphi_{k}} \sum_{n=0}^{\nu} a_{k-n} h_{\text{A},n} + n_{k}, 
\end{equation}
where \(y_{k}\) and \(\varphi_{k}\) denote the received sample and the sample of the time-varying PN process, respectively, at the \(k^{\text{th}}\) sample-time instance. The independent and identically distributed (i.i.d.) information sequence is denoted by \(a_{k}\), and it is drawn uniformly from the $M$-QAM constellation. \(h_{\text{A}}\) denotes the aggregate impulse response of the transmitter pulse shaping filter \(\hSSF\), dispersive channel \(h_{\text{C}}\), and receiver anti-aliasing filter \(h_{\text{R}}\) (see  Figure~\ref{TxAndRxInitDiagrams}), and it can be modelled as a causal filter of length \(\nu+1\). The signal at the input of the receiver is corrupted by noise samples $n_k$, which we assume to be additive white Gaussian noise (AWGN) with variance \(\sigma_{n}^{2}\) with the proper choice of the anti-aliasing filter \(h_{\text{R}}\).

The PN term in \eqref{DiscreteBasebandSystemEq} includes contributions from the transmitter and receiver LOs. Since for practical microwave systems PN is slowly time-varying with respect to the duration of the channel impulse response, this aggregation of the two PN processes into a single process is possible \cite{Falconer_PassbandDFE, Stark_CombiningDFEandCPE}. We adopt the commonly used Wiener model to characterize the PN process \cite{Colavolpe_IterativeDecodingPhaseNoise, Kreimer_EfficientIterativePhaseEstDecoding}, so 
\begin{equation}\label{PNasWienerProcess}
\varphi_{k} = \varphi_{k-1} + \psi_{k}, \enspace \psi_k \sim \mathcal{N}(0,\sigma_{\psi}^2),
\end{equation}
where $\mathcal{N}(0,\sigma_{\psi}^2)$ denotes the zero-mean normal distribution with differential PN variance $\sigma_{\psi}^2$. 

\subsection{Pulse Shaping Filter Design}\label{PulseShapingFilterDesign}

\begin{figure}[!t]
	\centering
	\includegraphics[width=1\columnwidth]{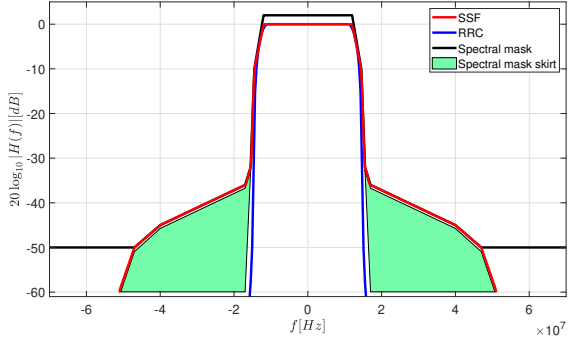}
	\caption{Magnitude response of the SSF and RRC pulse shaping filters, along with the seven-segment spectral mask, relative to the level at the carrier frequency.}
	\label{MagResp_SSF_RRC_Mask}
\end{figure}

Our proposed filter design aims at exploiting the spectrum skirts of regulatory emission masks for microwave systems. Figure~\ref{MagResp_SSF_RRC_Mask} (black line) shows an example of such a spectrum mask for transmission in the 17-30~GHz band as specified in \cite[Table 3e]{ETSISpectralMaskStandard}. The mask is a seven-segment envelope for the permitted PSD relative to the magnitude level at carrier frequency \(f_{\text{c}}\).

\textcolor{black}{Ideally, we would aim to design the SSF pulse shaping filter to maximize the AIR (as expressed in~\eqref{AIR}) as a measure for practically achievable rate, or, alternatively, minimize the mean squared error (MSE) of the equalized output at the receiver to account for equalization.
The latter approach has been pursued, for example, in \cite{Wen_TimeFreqCompressedFTN}, wherein pulse shaping involves a precoding filter applied to RRC pulses, and the receiver employs a matched filter followed by a minimum MSE (MMSE) linear equalizer (LE).
Then, a convex optimization problem is formulated to design the precoding filter and minimize the resulting MSE.
Different from \cite{Wen_TimeFreqCompressedFTN}, which is placed by the authors in the context of FtN transmission, we propose using the excess bandwidth under the mask skirts to enable faster signaling.
As it will be further discussed in Section~\ref{SimplifiedReceiverDSP}, such an SSF pulse causes severe ISI that is hard to compensate using linear equalization.
Thus, we resort to non-linear equalization which, unfortunately, renders the problem of transmit filter design for MSE minimization intractable and difficult to solve\footnote{\textcolor{black}{Unlike the MSE expression for LE utilized in~\cite{Wen_TimeFreqCompressedFTN}, which can be manipulated to put the optimization problem in a convex form, the MSE expression with decision-feedback equalization (DFE) involves pre- and post-multiplying the inverse of the regularized autocorrelation matrix with the DFE feedback coefficients (see, for example, \cite[(2.3.90)]{Fischer_Precoding}). Thus, all terms of the objective are functions of the design variable, i.e., SSF filter coefficients, which makes the problem difficult to handle.}}.
As a consequence, we opt to design the SSF filter to maximize the transmit energy under the mask, which is a fairly commonly used approach in filter design and leads to a standard convex optimization problem that can be efficiently solved.}

In particular, we adopt the frequency-domain approach introduced in \cite{Davidson_FIRFilterDesign} for the SSF pulse shape design. Our objective is to fill the spectral mask, which we formulate as the problem of minimizing the weighted sum squared error between the SSF filter frequency response \(\HSSF\) and the desired frequency response $\D$. \textcolor{black}{We write the design as the following convex optimization problem:
\begin{align}
&\underset{\hSSF}\minimize~~ \sum\limits_{i=1}^{K} w_i \left(|\D(f_i)| - |\HSSF(f_i)|\right)^2 \nonumber\\
&\text{subject to } \abs*{\HSSF(f_i)} \leq \abs*{\D(f_i)},\, i\!=\!1,\!2,\!\ldots\!,\!K,
\label{PS_WMSE}
\end{align}
where \(w_i \geq 0\) denote the weights, and the normalized frequency interval \([0, 1/2]\) is discretized using \textcolor{black}{\(K\)} points $f_i$, $i=1,2,\ldots, K$.}
\textcolor{black}{The real even-symmetrical frequency response can be written in matrix form as
\begin{align}
\label{FreqResp_matrix}
\HSSF\! 
&=\!
\begingroup 
\setlength\arraycolsep{2pt}
 \begin{pmatrix}
  1 & 2 & \cdots & 2 \\
  1 & 2\cos(\frac{2\pi}{K}) & \cdots & 2\cos(\frac{2\pi M}{K}) \\
  \vdots  & \vdots  & \ddots & \vdots  \\
  1 & 2\cos(\frac{2\pi (K-1)}{K})\! & \cdots & \!2\cos(\frac{2\pi (K-1)M}{K})
 \end{pmatrix}\!\!\!\!
 \endgroup
 \begin{pmatrix}
   h_0 \\
   h_1 \\
   \vdots \\
   h_M
  \end{pmatrix} \nonumber \\
  &\equiv \V \hSSFv,
\end{align}
where \(\hSSFv\) is half of the SSF response.} 
\textcolor{black}{We rewrite the design problem in \eqref{PS_WMSE} in matrix form, using \eqref{FreqResp_matrix}, as
\begin{align}
\label{min_errorNorm}
&\underset{\hSSFv}\minimize ~~ \norm{\w^{1/2} \odot \left(\dT - \V \hSSFv\right)}^2 \nonumber \\
&\st ~~ \left(\V \hSSFv\right)^{\abs{\cdot}} \leq \dT,
\end{align}
where \(\dT = [|\D(f_1)|,|\D(f_2)|,\ldots,|\D(f_K)|]^\text{T}\) and $\w=[w_1,w_2,\ldots,w_K]^\text{T}$.
The problem in \eqref{min_errorNorm} is convex quadratic and 
can be solved in polynomial time in $K$ via ellipsoid or interior point algorithms. 
}

Figure~\ref{MagResp_SSF_RRC_Mask} shows the magnitude response of a designed SSF pulse shaping filter (red line) for the case of $N_{\hSSF} = 257$ filter coefficients and a skirt ranging from \(15.5 \text{~MHz}\) to \(51.2\text{~MHz}\), corresponding to a power level ranging from $-32$~dB to $-60$~dB relative to $0$~dB at the carrier frequency.
The figure also includes the commonly used RRC filter and the adopted seven-segment spectral mask. The spectral skirts, i.e., the used guard bands, are highlighted in the figure. We observe how the SSF filter permits signal transmission in the spectral skirts up to the mask limit. 
\textcolor{black}{In particular, since the SNR experienced by fixed microwave transmission is often greater than 50~dB, the energy of the pulse in the spectrum skirts will be adequate so as to support an overall higher rate through wideband transmission.}
For example, comparing Nyquist transmission with $\Rrrc=25.6$~Msymbol/s for the RRC filter, and wideband signaling with $\Rssf=51.2$~Msymbol/s for the SSF filter in Figure~\ref{MagResp_SSF_RRC_Mask}, assuming an SNR of 50~dB at the carrier frequency and a maximal constellation size of $M=4096$, the communication-theoretic rate limit  \cite[Eq.~(5)]{Anderson_FTNSignaling} is increased from 307~Mbit/s to 455~Mbit/s, i.e., an improvement of 50\% is possible.

\subsection{Equalization and Precoding}\label{SimplifiedReceiverDSP}

\begin{figure}[!t]
	\centering
	\includegraphics[width=1\columnwidth]{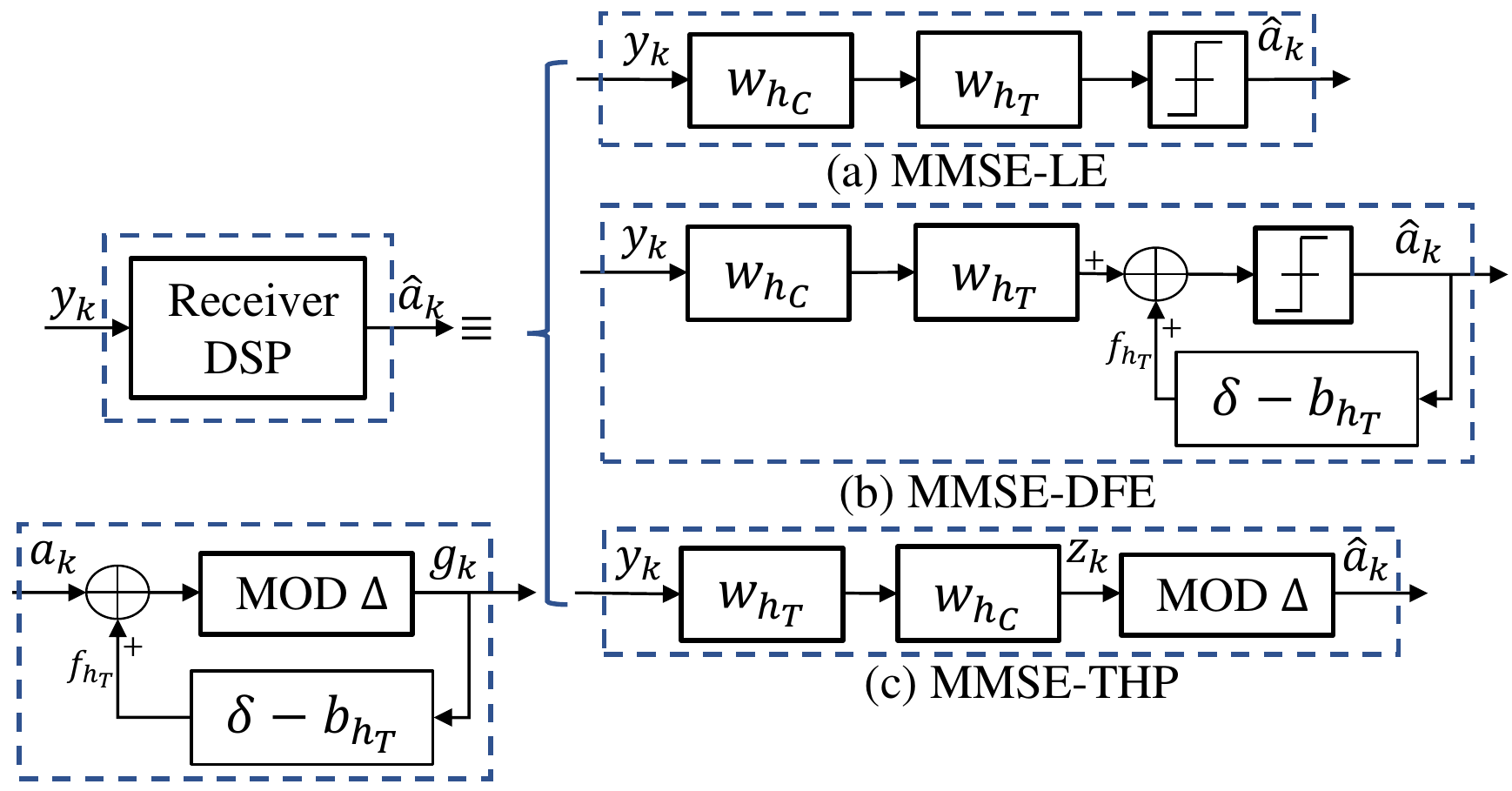}
	\caption{Three structures of equalization and pre-equalization for combating the SSF filter generated ISI: (a) MMSE-LE; (b) MMSE-DFE; (c) MMSE-THP. The dispersive channel ISI is mitigated via an additional MMSE-LE filter.}
	\label{SimplifiedRxDSPDiagram}
\end{figure}

Conventional microwave systems employ adaptive linear equalization to mitigate the typically mild ISI caused by multipath propagation \cite{Qureshi_AdaptiveEqualization}. Considering an SSF design as illustrated in Figure~\ref{MagResp_SSF_RRC_Mask}, we expect that further equalization is required to compensate for the ISI induced by pulse shaping. However, since the SSF filter is known \textsl{a priori}, non-adaptive equalization with pre-computed filters can be applied. This structure is illustrated in Figure~\ref{SimplifiedRxDSPDiagram}(a), where the received signal is processed through an LE filter $w_{h_{\text{C}}}$ for the channel-induced ISI, followed by another LE filter $w_{h_\text{T}}$ for the ISI due to pulse shaping. For both filters, we apply the MMSE design criterion, but only the relatively short filter $w_{h_\text{C}}$ is adapted using training signals, as it is done in conventional microwave systems.

Since the ISI from pulse shaping is severe, more powerful equalization techniques need to be used. Figure~\ref{SimplifiedRxDSPDiagram}(b) shows the application of DFE \cite{AlDhahir_MMSEDFE} with the feedforward filter (FFF) $w_{h_\text{T}}$ and feedback filter (FBF) $b_{h_\text{T}}$ in tandem with the adaptive LE $w_{h_{\text{C}}}$. 
\textcolor{black}{An MMSE-DFE has the advantage of not enhancing the noise, as opposed to an LE, and it mitigates ISI stemming from channels with deep notches.}
Finally, THP \cite{Tomlinson_THP, Harashima_THP}, i.e., non-linear pre-equalization at the transmitter, is well suited for the equalization of the SSF-induced ISI, since off-line computation of the filters avoids the need for adaptation through feedback from the receiver to the transmitter. This structure is shown in Figure~\ref{SimplifiedRxDSPDiagram}(c), which includes the FBF $b_{h_\text{T}}$ and the modulo operation
\begin{equation}
\label{eq:mod}
\textnormal{MOD}_{\Delta} (x) = x - \Delta\ \round{\frac{\Re\{x\}}{\Delta}} -\ju \Delta \round{\frac{\Im\{x\}}{\Delta}}
\end{equation}
at the transmitter, and the FFF $w_{h_\text{T}}$ along with modulo operation at the receiver. In \eqref{eq:mod}, \(\Delta=\sqrt{M} d_{\textnormal{min}}\) denotes the modulo constant, where \(d_{\textnormal{min}}\) is the minimum Euclidean intersymbol distance. Compared to DFE, THP has the advantage of zero error propagation.

As we will show through numerical results in Section~\ref{NumericalResultsDiscussion}, THP is the method of choice for equalization of the proposed wideband transmission. In the following two sections, we present effective designs to combine THP with phase tracking and estimation for PN compensation, as PN is the key performance limiting impairment for microwave transmission systems.

\section{Nonlinear Precoding and DPLL Phase Noise Tracking}\label{NonlinearPrecoding_DPLL_CPE}

Phase noise limits the permissible constellation size and thus the achievable spectral efficiency of microwave systems. 
This is due to higher sensitivity of larger signal constellations to PN. 
Therefore, in this section, we combine THP for equalization as described above with a DPLL for PN compensation. We refer to this approach as THP-DPLL. As we will elaborate on, the challenge of THP-DPLL lies in the operation on the extended signal constellation created by THP at the receiver.

\subsection{The Phase-Tracking Loop}


We consider the discrete-time baseband signal model in \eqref{DiscreteBasebandSystemEq}, and ignore, for the moment, the presence of ISI. 
\textcolor{black}{Let $\hat{\varphi}_{k}$ and $\hat{a}_{k}$ denote the receiver estimates of the phase error $\varphi_{k}$ and \(k^{\text{th}}\) transmitted symbol $a_{k}$, respectively. Then, a first-order DPLL applies the update}
\begin{equation}
	\hat{\varphi}_{k+1} = \hat{\varphi}_{k} + G e_{k} \label{DPLL_Eq},
\end{equation}
where $G$ is the loop gain and the error signal $e_{k}$ is obtained from \cite[Eq.~(8)]{Stark_CombiningDFEandCPE}
\begin{equation}
e_{k} = \Im\left\{y_{k} \left(\hat{a}_{k} \e^{\ju \hat{\varphi}_{k}} \right)^*\right\}\label{error_term}.
\end{equation}
\textcolor{black}{Tracking the phase involves computing the error signal as in \eqref{error_term} and updating the phase error estimate according to \eqref{DPLL_Eq}.}

\subsection{Combining THP and the DPLL}

\begin{figure}[!t]
	\centering
	\includegraphics[width=1 \columnwidth]{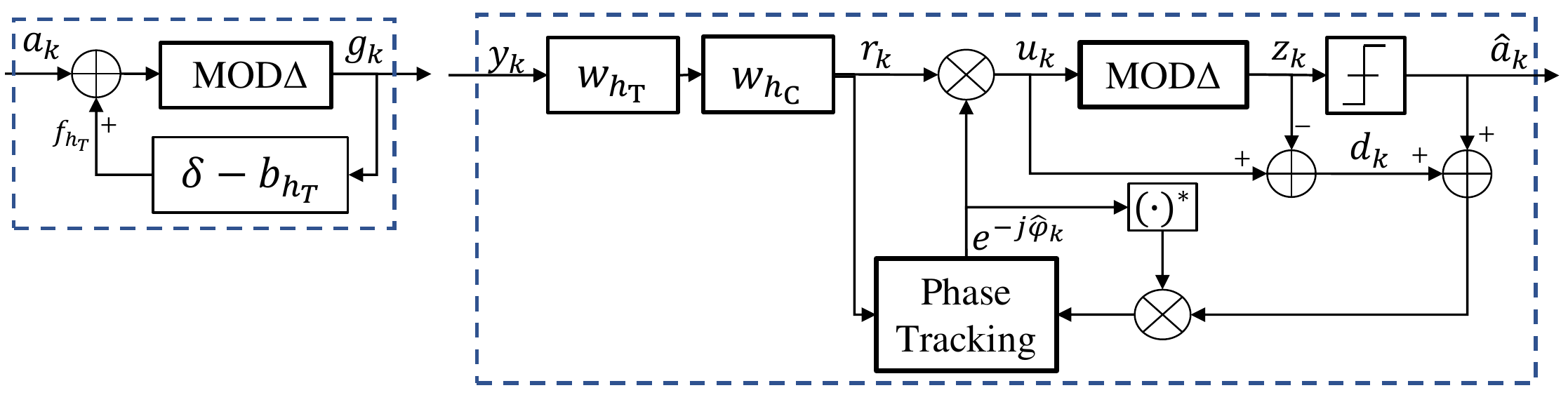}
	\caption{The proposed THP-DPLL transceiver configuration. }
	\label{THP_DPLL_Tx_Rx}
\end{figure}

We perform joint precoding and PN compensation through THP-DPLL, as illustrated in Figure~\ref{THP_DPLL_Tx_Rx}. At receiver, phase correction takes place after the THP FFF, but before the modulo device. This is required because a phase rotation and the modulo operation are not interchangeable. Thus, PN compensation has to be applied to the THP extended signal constellation. 

\subsubsection{Phase Tracking}

Phase tracking is based on the computation of the error signal in \eqref{error_term} using estimates of the transmitted symbol $\hat{a}_{k}$. To map this estimate to an estimate in the THP extended constellation, we add the displacement term $\hat{d}_{k} = u_{k} - \textnormal{MOD}_{\Delta} (u_{k})$ as shown Figure~\ref{THP_DPLL_Tx_Rx}. Hence, the error term from \eqref{error_term} can be rewritten as
\begin{equation} 
e_{k} = \Im\left\{u_{k} \left((\hat{a}_{k}+\hat{d}_{k}) \e^{\ju \hat{\varphi}_{k}} \right)^*\right\}.
\label{error_term_extended}
\end{equation}

\subsubsection{Pilot Symbols}
The use of pilot symbols, i.e., replacing the decisions $\hat{a}_k$ in \eqref{error_term_extended} with a reference signal $a^{\text{pilot}}_k$, is essential for initializing the operation of the DPLL. Furthermore, pilots are necessary throughout the transmission to help the DPLL maintain its lock to the carrier phase. In the context of THP transmission, we note that the effectiveness of such pilot-aided carrier phase recovery depends on the accuracy of the reference signal $a^{\text{pilot}}_k+\hat{d}_k$ in the extended constellation. The reference signal is the estimate of the representation $u_k^{\text{pilot}}=a^{\text{pilot}}_k+d_k$ of the pilot signal $a^{\text{pilot}}_k$ in the extended constellation. This means that even in the instance of pilot-based training, the reference signal can be erroneous, since $\hat{d}_k\neq d_k$ is possible. We note that $d_k$ depends on data symbols preceding the pilot, and thus it is not known \textsl{a priori} at the receiver.

The decision on \(d_k\) reduces to correctly identifying the region in which the pilot signal representation $u^{\text{pilot}}_k$ is located. By expressing \(d_k\) as \(d_k = \mu_{k} \Delta + \ju \nu_{k} \Delta\), and denoting the output of the THP FFF as \(r_k\), the decision on \(d_k\), and subsequently on \(u^{\text{pilot}}_k\), can be performed as
\begin{align}
\label{e:dkDecision}
\hat{\mu}_{k} &=\round{\Re\{r_{k} e^{-\ju \hat{\varphi}_{k}} - a^{\text{pilot}}_k\} / \Delta}, \nonumber \\
\hat{\nu}_{k} &=\round{\Im\{r_{k} e^{-\ju \hat{\varphi}_{k}} - a^{\text{pilot}}_k\} / \Delta},\\
\label{e:ukHat}
\hat{u}^{\text{pilot}}_k &=  a^{\text{pilot}}_k + \hat{\mu}_{k} \Delta + \ju \hat{\nu}_{k} \Delta.
\end{align}

\section{Nonlinear Precoding and BCJR Phase Noise Estimation}\label{NonlinearPrecoding_BCJR_CPE}

Bayesian methods make use of the model for the underlying stochastic process, and thus they are typically more effective in combating PN compared to phase-tracking loops.
In this section, we propose a joint Bayesian PN estimation approach in tandem with THP. Specifically, we combine THP and the BCJR sequence estimator, and refer to this proposed approach as THP-BCJR. Again, the operation in the extended signal space requires particular attention.

\subsection{BCJR-based Phase Noise Estimation}
The BCJR algorithm \cite{BCJR_OptimalDecoding} performs inference for a hidden Markov process, and it can naturally be applied to the PN model \eqref{PNasWienerProcess}, either in the form of message passing of parameters of distribution approximations \cite{Colavolpe_IterativeDecodingPhaseNoise} or on a quantized phase trellis \cite{Ferrari_UnifiedFrameworkDetection}. We use the latter approach. In particular, the pilot-assisted PN estimation presented in \cite{Hoda_AccurateBCJRbasedCPE}, which is well-suited for high-SNR scenarios, serves as a \textcolor{black}{basis for our method. Since \cite{Hoda_AccurateBCJRbasedCPE} uses orthogonal transmission with RRC pulses, it starts from the assumption of reliable PN estimation samples at pilot positions.} 

The phase trellis is built between each two consecutive pilot symbols, such that it starts and ends in a single state. The state space is defined by the maximal PN span interval $\Iphi=[-\phimax, \phimax]$ with respect to the phase at the pilot at the beginning of the block, and the number $\Sphi$ of discretized phases so that $\varphi_k \in \{-\phimax+2k\phimax/\Sphi, \,k=0,1,\ldots,\Sphi-1\}$. The choice of $\phimax$ depends on pilot spacing $d_{\text{pilot}}$ and PN variance $\sigma_{\psi}^2$. For example, by setting $\phimax=3\sqrt{\sigma_\psi^2d_{\text{pilot}}}$, the probability of the actual phase at the end of the current block falling outside of $\Iphi$ is less than 0.3\%. The phase-quantization interval should be sufficiently small to have negligible effect given a constellation size $M$. On the other hand, the number of phase levels $\Sphi$ should be selected conservatively as it determines the size of the trellis BCJR operates on and thus the complexity of the PN estimation. If this becomes an issue, per-state survivor processing as described in \cite{Colavolpe_ReducedStateBCJR} can be applied to control the size of the BCJR state space.

The transition probabilities from phase state $\varphi'$ to $\varphi$ used in BCJR PN estimation are given by
\begin{equation}\label{PN_transition_probabilities}
\gamma_k(\varphi', \varphi)\! = \!p_\psi(\psi\! = \!\varphi'\! - \!\varphi) \, p_n(n\! = \!r_k \e^{-\ju \varphi}\! - \!\hat{a}_k)\;,
\end{equation}
where $p_\psi$ is the distribution of the discretized PN increment, $p_n$ is the distribution of the AWGN, and $r_k$ and $\hat{a}_k$ are the received sample input to the BCJR-based PN estimation and the tentative symbol decision associated with the PN-state $\varphi$, respectively. The final PN estimate is the phase associated with the most likely trellis state for each time step. The choice of $\hat{a}_k$ for THP transmission will be discussed in the following.

\subsection{Joint Precoding and BCJR-based Phase Noise Estimation}

\begin{figure}[!t]
	\centering
	\includegraphics[width=1\columnwidth]{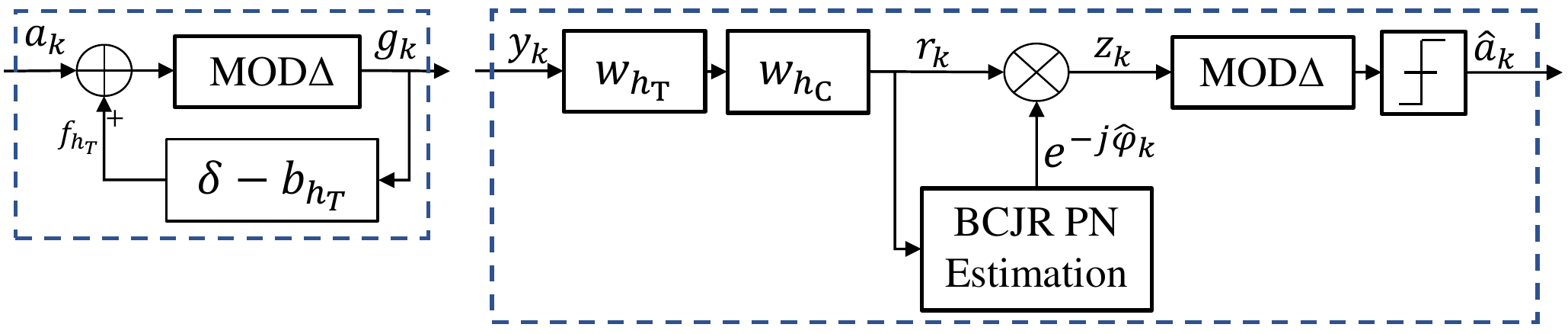}
	\caption{THP and BCJR-based PN estimation system model.}
	\label{THP_SSF_BCJR_completeDiagram}
\end{figure}

Figure~\ref{THP_SSF_BCJR_completeDiagram} depicts the proposed joint precoding and BCJR-based PN estimation.
Similar to THP-DPLL introduced in Section~\ref{NonlinearPrecoding_DPLL_CPE}, we need to place the PN compensation between the FFF and the modulo device. Hence, we need to adjust the baseline BCJR-based PN estimation described above to work on the extended constellation space. 

\subsubsection{THP-BCJR}
In case of THP, the term $r_k \e^{-\ju \varphi} - \hat{a}_k$ in \eqref{PN_transition_probabilities} would be replaced by $r_k \e^{-\ju \varphi} - \hat{u}_k$, where $\hat{u}_k$ is the tentative decision of the transmitted signal from the THP extended constellation, i.e., the signal point in the extended constellation closest to $r_k \e^{-\ju \varphi}$. However, since   
\begin{equation}
|r_k \e^{-\ju \varphi} - \hat{u}_k|\le d_{\textnormal{min}}/2,
\end{equation}
we directly integrate the modulo operation into the computation of the branch metric as
\begin{align}
\gamma_k(\varphi', \varphi) &= p_{\psi_k}(\psi_k = \varphi'-\varphi) \nonumber \\
&\times p_n\left(n\! = \!\textnormal{MOD}_{\Delta}\left(r_k \e^{-\ju \varphi}\right)- \hat{a}_k\right)\;,
\end{align}
where $\hat{a}_k$ is the data symbol estimate on the non-extended constellation assuming the PN is $\varphi$.

The output of the BCJR PN estimation is then used to perform PN compensation on the extended constellation symbols, followed by the modulo operation to fold the PN-compensated symbols back into the original QAM constellation (see Figure~\ref{THP_SSF_BCJR_completeDiagram}).

\subsubsection{Pilot Symbols}

The pilot-assisted BCJR PN estimation \cite{Hoda_AccurateBCJRbasedCPE} relies on initial phase estimates for both the forward and backward recursions. Similar to the case of THP-DPLL, THP-BCJR also faces the problem of constellation extension due to precoding. 

\begin{figure}[!t]
	\centering
	\subfloat[Empirical PMF of effective pilots.]{%
		\includegraphics[width=1\columnwidth]{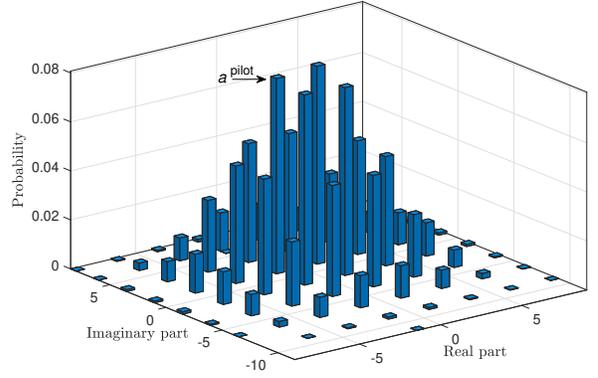}}
		\hfill
	\subfloat[Empirical PMF of the magnitude of the effective pilots.]{%
		\includegraphics[width=1\columnwidth]{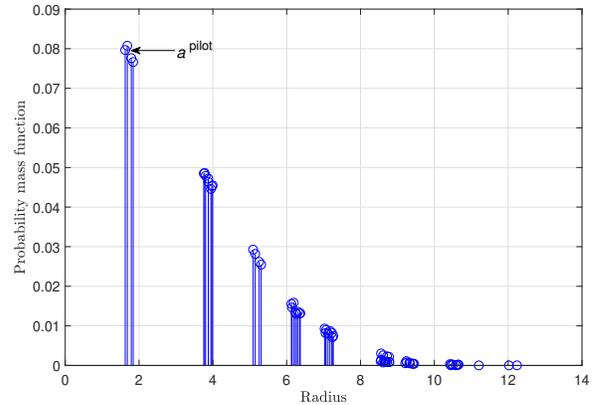}}
\caption{Empirical PMF of the impairment-free effective pilots, corresponding to a transmitted pilot $a_k^{\text{pilot}} = -31/\sqrt{682} + \ju 29/\sqrt{682}$ from a unit-norm $1024$-QAM constellation, and an SSF design as illustrated in Figure~\ref{MagResp_SSF_RRC_Mask}.}
	\label{ReferencePilots_PMF}
\end{figure}

First, let us consider a pilot symbol $a^{\text{pilot}}_k$, which is known at the receiver. Due to THP, the effective pilot symbol $u^{\text{pilot}}_k = a^{\text{pilot}}_k + d_k$, $d_k = (\ell_k+\ju m_k)\Delta$, $\ell_k, m_k \in \mathbb{Z}$, is a random variable. Figure~\ref{ReferencePilots_PMF}(a) shows the empirical probability mass function (PMF) of the impairment-free effective pilot $u_k^{\text{pilot}}$ when transmitting the pilot symbol $a_k^{\text{pilot}} = -31/\sqrt{682} + \ju 29/\sqrt{682}$ from a unit-norm \(1024\text{-QAM}\) constellation, and an SSF design as illustrated in Figure~\ref{MagResp_SSF_RRC_Mask}. Observing the ambiguity in the effective pilot symbol, we formulate the PN estimation as a joint effective pilot and PN search task, making use of the \textsl{a priori} probabilities for $u^{\text{pilot}}_{k}$
\begin{align}
\label{e:pilotjoint}
\left(\hat{\varphi}_k,\hat{u}^{\text{pilot}}_{k}\right) &= \argmax\limits_{\varphi\in[0,2\pi), u \in {\cal U}}\Big\{-\left|r_{k} - u\e^{\ju \varphi}\right|^2/\sigma_n^2 \nonumber \\
&+ \log\left(\Pr(u^{\text{pilot}}_{k} = u)\right)\Big\}\;,
\end{align}
where ${\cal U}=\{a^{\text{pilot}}_{k} + \Delta \mathbb{Z}^2\}$ is the set of effective pilot symbols. The estimation in \eqref{e:pilotjoint} can also be written as
\begin{align}
\label{e:pilotGRLT}
\hat{u}^{\text{pilot}}_{k} =& \argmax\limits_{u \in {\cal U}}\Big\{-\left(|r_{k}|-|u|\right)^2/\sigma_n^2 \nonumber \\
+& \log\left(\Pr(u^{\text{pilot}}_{k} = u)\right)\Big\}\\
\label{e:pilotGRLT2}
\hat{\varphi}_k=&\arg\left\{r_{k}\left(\hat{u}_{k}^{\text{pilot}}\right)^*\right\}\;.
\end{align}
We observe from \eqref{e:pilotGRLT} that only the magnitude of the pilot symbols in the extended constellation matters, which, together with similar \textsl{a priori} probabilities, renders the joint estimation as per \eqref{e:pilotjoint} error prone.  Figure~\ref{ReferencePilots_PMF}(b) illustrates this for the scenario shown in Figure~\ref{ReferencePilots_PMF}(a), and we note the small distances between different magnitudes of possible effective pilot symbols. In the following, we propose a solution to obtain reliable initial phase estimates for the BCJR PN estimation. 

\paragraph{Improving Initial Phase Estimates}
Similar to THP-DPLL, we use coarse phase estimates $\hat\varphi_{\text{coarse}}$ to support the estimation of the effective pilot symbol.

\begin{figure}[t]
	\centering
	\subfloat[Projections of all effective pilots $u_k\in{\cal U}$ on the unit circle.]{%
		\includegraphics[width=0.8\columnwidth]{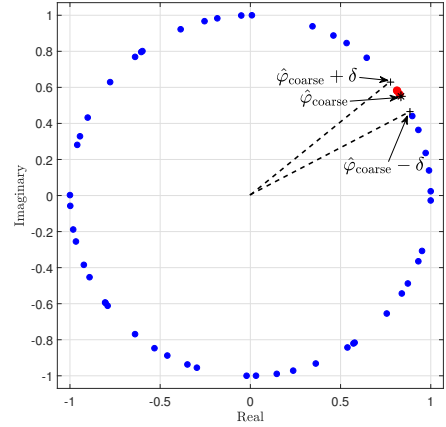}}
		\hfill
	\subfloat[Empirical PMF of the magnitude of the effective pilots within the admissible phase range.]{%
		\includegraphics[width=0.8\columnwidth]{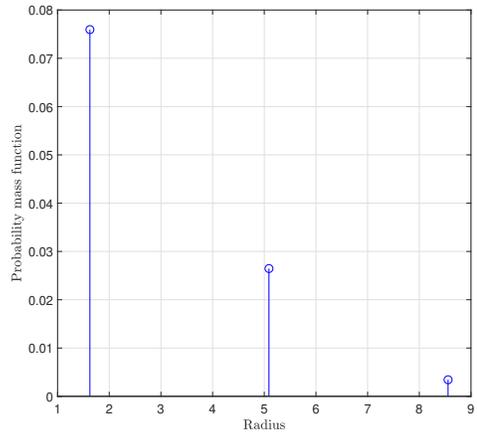}}
\caption{Use of a coarse PN estimate $\hat\varphi_{\text{coarse}}$ and the projections of the effective pilots $u_k \in {\cal U}$ to reduce the range of admissible PN values; parameters correspond to Figure~\ref{ReferencePilots_PMF}, and $\hat\varphi_{\text{coarse}} = 0.58$~rad, and $\delta = 0.1$~rad.}
	\label{PNSearchSpaceLimit}
\end{figure}

Given $\hat\varphi_{\text{coarse}}$, we limit the phase search space to ${\cal F}=\{\hat\varphi_{\text{coarse}}+[-\delta,\delta]\}$, with parameter $0<\delta<\pi$, in  \eqref{e:pilotGRLT}. Figure~\ref{PNSearchSpaceLimit}(a) illustrates the effect of limiting the range of admissible PN values by showing the projections of all $u_k\in{\cal U}$ onto the unit circle for the example from Figure~\ref{ReferencePilots_PMF}, for the case of $\hat\varphi_{\text{coarse}} = 0.58$~rad and $\delta = 0.1$~rad. For the effective pilots within the admissible phase range, the metric in \eqref{e:pilotGRLT} remains unchanged. The search space restriction largely discourages the ambiguities in the effective pilot symbols identified in Figure~\ref{ReferencePilots_PMF}(b). This is illustrated in Figure~\ref{PNSearchSpaceLimit}(b), which only shows the effective pilots which fall into the admissible phase range.

An extreme case is to reduce the search space to the coarse phase estimate itself, i.e., ${\cal F}=\{\hat\varphi_{\text{coarse}}\}$. Then we have
\begin{align}
\label{e:CoarsePNEstSearchSpace}
\hat{u}^{\text{pilot}}_{k} &= \argmax\limits_{u\in {\cal U}}\Big\{-\left|r_{k} - u \e^{\ju \hat\varphi_{\text{coarse}}}\right|^2/\sigma_n^2 \nonumber \\
&+ \log\left(\Pr(u^{\text{pilot}}_{k} = u)\right)\Big\}\;.
\end{align}
Given the estimated pilot, the refined PN estimate $\hat\varphi_k$ is obtained from \eqref{e:pilotGRLT2}.

\paragraph{Coarse Phase Estimates}
To obtain coarse phase estimates, we modify the block-based structure adopted from \cite{Hoda_AccurateBCJRbasedCPE} for BCJR PN estimation, which considered independent PN estimation for the blocks of data symbols located between two pilot symbols. In particular, we use an extrapolation of PN estimates from the previous block as the coarse phase estimate for the pilot symbol at the beginning of the current block of data symbols. However, this is not possible for the pilot location at the end of the current block due to the phase changes during a block duration. Instead, we use the best tentative phase estimate at the end of the forward recursion as the coarse phase estimate to obtain the effective pilot symbol for the backward recursion. 

\section{Numerical Results and Discussion}\label{NumericalResultsDiscussion}

In this section, we assess the performance of the proposed wideband transmission and PN compensation methods via software simulations of a typical microwave backhaul link.

\subsection{Microwave Backhaul Transmission Setup}

We consider point-to-point microwave transmission with the structure shown in Figure~\ref{TxAndRxInitDiagrams} and the system and channel parameters corresponding to the specifications of a typical backhaul link. The conventional RRC pulse shape with a roll-off factor $\beta = 0.15$ is applied for the benchmark system. The symbol rate is set to $R_{\text{RRC}} = 25.6$~Msymbol/s, which leads to a signal bandwidth of $B_{\text{RRC}} = 29.4$~MHz. This is suitable for transmission in the 17-30~GHz band using the seven-segment spectrum mask as specified in \cite[Table 3e]{ETSISpectralMaskStandard}. For the proposed system using the spectrum skirts, we design an SSF filter with $N_{\hSSF} = 257$ coefficients fitting the mask as shown in Figure~\ref{MagResp_SSF_RRC_Mask}. The design bandwidth, \textcolor{black}{including the skirts, extends to} $B_{\text{SSF}} = 4R_{\text{RRC}} = 102.4$~MHz, and we obtain the SSF filter by discretizing the normalized frequency range $[0,0.5]$ into 1,000 equally sized segments and setting the weight vectors $\w^{1/2}(\{f|0\le f < f_4\})=1$, $\w^{1/2}(\{f|f_4\le f<f_5\}) = 10$, and $\w^{1/2}(\{f|f_5\le f\le 0.5\}) = 1000$ for the optimization in \textcolor{black}{\eqref{min_errorNorm}}, where $[f_1,f_2,\ldots,f_7]=[0.11, 0.14, 0.15, 0.16, 0.39, 0.46, 0.5]$ are the corner frequencies of the seven-segment mask. 
\textcolor{black}{The SSF-based transmission is viable only when the neighbouring channels are not active.
But, in the unlikely event that a neighbouring channel is active, we will stop shaping the signal with an SSF, and revert to using an RRC filter, which limits the signaling rate to a half of the one we could use with an SSF.}
The variances of the transmitter and receiver LO PN processes are set according to a PN spectral density $\mathcal{L}(f) = -90$~dBc/Hz at an offset frequency $f_{\text{offset}} = 100$~kHz from the carrier frequency, which is a typical level for practical microwave systems \cite{Kreimer_EfficientIterativePhaseEstDecoding}. This corresponds to a PN variance $\sigma_\psi^2=\textcolor{black}{10^{\mathcal{L}(f)/10} 4\pi^2 f^2_{\text{offset}} T_{\text{RRC}} =}1.5 \times 10^{-5} $~rad$^2$ for the symbol period $T_{\text{RRC}}\approx 39$~ns for the RRC-based transmission. 
The signal propagation model is represented by a Rummler channel with the distributions for the notch frequency and notch depth taken from \cite{Rummler_MultipathFading}. The results shown in the following are the performances averaged over 1,000 Rummler channel realizations. The receiver input anti-aliasing filter is an RRC filter with the bandwidth set according to the RRC or SSF transmission rate, resulting in white noise after sampling, and a roll-off factor \(\beta = 0.15\) is chosen. 

\subsection{Performance Measures}

For assessing the performance of the proposed SSF scheme, we consider the SER of uncoded transmission and the AIR \textcolor{black}{and BER} for coded transmission. The AIR can be obtained from \cite[Eq.~(6)]{Eriksson_AchievableRates}
	\begin{align}
	\text{AIR} &= m  - \frac{1}{N} \sum_{i = 1}^{m} \sum_{k = 1}^{N} \log_2 \Big(1 \nonumber \\
	&+ \exp\left((-1)^{b_{k,i}}\mathrm{LLR}_{k,i}\right) \Big),
	\label{AIR}
	\end{align}
where $m=\log_2(M)$, $N$ is the length of the transmitted bit sequence, \(b_{k,i}\in\{0,1\}\) is the binary value and \(\mathrm{LLR}_{k,i}\) is the  log-likelihood ratio (LLR) for position $i$ in the label of the $k$th transmitted symbol, respectively.

We compare systems with the proposed SSF and the conventional RRC pulse shaping meeting the spectral mask constraints. This leads to slightly different SNRs, since the transmit power for the SSF-based transmission is marginally higher due to better filling of the spectral mask. For the following, \textcolor{black}{we define the SNR as the symbol energy divided by the noise power per bandwidth unit (or PSD), that is, \(E_{\text{s}}/N_0\), of the baseline RRC-shaped transmission.}
We use the SNR for the RRC-based system when specifying performance as a function of SNR. 

\subsection{Numerical Results}

\begin{table}
\begin{center}
\caption{Maximum data rate for conventional RRC and the SSF pulse shape system with different QAM constellation sizes.} \label{tab:BitRateVsConstSize}
\begin{tabular}{ccc}
\toprule
	& QAM order & QAM order \\
	Maximum data rate \(\left[\text{Mbit/s}\right]\) & RRC	& SSF \\
\midrule
	\(307\)	& \(2^{12}\) & \(2^{6}\) \\
\midrule
	\(358\) & \(2^{14}\) & \(2^{7}\) \\
\midrule
	\(410\)	& \(2^{16}\) & \(2^{8}\) \\
\midrule
	\(461\) & \(2^{18}\) & \(2^{9}\) \\
\midrule
	\(512\)	& \(2^{20}\) & \(2^{10}\) \\
\bottomrule
\end{tabular}
\end{center}
\end{table}

We operate the system with the SSF pulse shaping filter at a rate of $R_{\text{SSF}} = 2R_{\text{RRC}} = 51.2$~Msymbol/s throughout the numerical results. This enables a significant reduction in constellation size for achieving the same maximal rate as the conventional RRC system with $R_{\text{RRC}} = 25.6$~Msymbol/s, while still permitting effective (pre-)equalization of the pulse-shaped induced ISI. Table~\ref{tab:BitRateVsConstSize} \textcolor{black}{highlights the rates that could be achieved by both systems for various QAM constellation sizes $M$.}
For example, to double the maximal data rate from 256~Mbit/s to 512~Mbit/s in the conventional RRC system, the QAM constellation size needs to increase to $M=2^{20}$, for which reliable communication is challenging due to PN distortion. The proposed system with SSF pulse shaping can achieve 512~Mbit/s with only $M=2^{10}$, while meeting the same spectral mask requirements as the RRC system. In the following, we will present results \textcolor{black}{in terms of AIR, pre-FEC SER, and post-FEC BER} that demonstrate to what extent SSF based transmission can meet expectations when using the practical equalization and PN compensation methods introduced above.

\subsubsection{Comparison of Equalization and Pre-equalization Methods}

\begin{figure}[t]
	\centering
	\includegraphics[width=1\columnwidth]{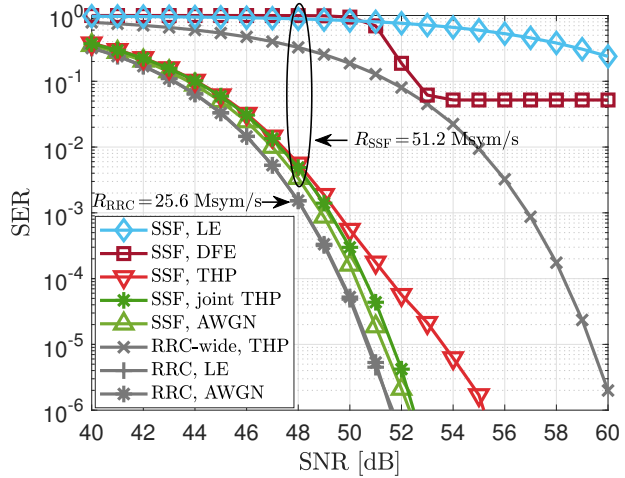}
	\caption{\textcolor{black}{SER vs.\ SNR for SSF pulse-shaped transmission with MMSE-LE (``LE''), MMSE-DFE (``DFE'') and MMSE-THP (``THP''). QAM constellation size $M=2^{10}$ and ideal carrier generation are assumed. We include reference curves for (i) MMSE-THP for the joint SSF pulse shape and Rummler channel (``SSF, joint THP''), (ii) SSF-based transmission in a non-dispersive AWGN channel (``SSF, AWGN''), (iii)  THP and SSF designed as a shortened RRC pulse shaping filter (``RRC-wide, THP''), (iv) the conventional RRC-based transmission with $M=2^{14}$ (``RRC, LE''), and (v) the conventional RRC-based transmission with $M=2^{14}$ in a non-dispersive AWGN channel (``RRC, AWGN'').}}
	\label{SER_MMSELE_MMSEDFE_MMSETHP_AWGN_Rummler_noTxRxImpairm}
\end{figure}

We analyse the performance achieved with MMSE-LE, MMSE-DFE and MMSE-THP introduced in Section~\ref{SimplifiedReceiverDSP} and illustrated in Figure~\ref{SimplifiedRxDSPDiagram}. For this purpose, we assume the operation of the microwave link with an ideal carrier phase generation. Figure~\ref{SER_MMSELE_MMSEDFE_MMSETHP_AWGN_Rummler_noTxRxImpairm} shows the SER achieved by the shaped wideband transmission with $M=2^{10}$ QAM over Rummler channels. As reference curves, we include
(i) the SER for MMSE-THP designed to jointly equalize the ISI due to SSF pulse shaping and Rummler channel (``SSF, joint THP''), 
\textcolor{black}{(ii) the SER for the SSF transmission over a non-dispersive AWGN channel (``SSF, AWGN'')},
(iii) the SER for wideband ($\Rssf=51.2$~Msymbol/s) transmission using a truncated\textcolor{black}{\footnote{\color{black} We truncate the RRC filter to 16 taps in order to transmit signal energy in the spectrum skirts.}} RRC filter that still meets the spectrum mask constraints (``RRC-wide, THP''),
(iv) the SER for conventional RRC transmission with $M=2^{14}$ (``RRC, LE''), and
\textcolor{black}{(v) the SER for the conventional RRC transmission with $M=2^{14}$ which propagates through a non-dispersive AWGN channel (``RRC, AWGN'').}

We observe that MMSE-LE is not able to equalize the SSF-induced ISI for the SNR range of interest.
\textcolor{black}{However, for the RRC-shaped transmission with $M=2^{14}$, an MMSE-LE effectively mitigates the ISI, which stems only from the dispersive Rummler channel.}
MMSE-DFE manages to lower the SER at an SNR of around 50~dB, \textcolor{black}{but it experiences an error floor of \(5 \times 10^{-2}\)}. Furthermore, there is a large gap to the performance achieved with MMSE-THP, which effectively combats the pulse shaping ISI.
This suggests that error propagation is the reason for the relatively poor performance of MMSE-DFE. \textcolor{black}{The large taps magnitude of the FBF of MMSE-DFE for the chosen SSF pulse shape plays a role in propagating decision errors.}
We conclude that MMSE-THP is the method of choice for equalization in SSF-based transmission.

The comparison of the SER results for THP and joint THP indicates that the LE for the Rummler channel ISI is not limiting the performance considering typical pre-FEC target SERs in the range of $10^{-2}$. Hence, the combination of THP at the transmitter for the static SSF pulse shape and adaptive LE at the receiver for the slowly time-varying channel as shown in Figure~\ref{SimplifiedRxDSPDiagram}(c) is a practical and performance-efficient solution.

To assess the effectiveness of the proposed SSF pulse shaping design presented in Section~\ref{PulseShapingFilterDesign}, we consider wideband transmission using a truncated RRC filter as SSF pulse shape, i.e., the RRC filter side-lobes in the skirts are used for transmission. We observe that the designed SSF pulse shape provides more than 6~dB gain over the truncated RRC filter, when applying THP to both. This clearly justifies the SSF filter design proposed in this paper. 
  
Finally, we observe from Figure~\ref{SER_MMSELE_MMSEDFE_MMSETHP_AWGN_Rummler_noTxRxImpairm} that conventional RRC-based transmission with a QAM size of $M=2^{14}$ \textcolor{black}{has a minimal SNR gain for a target SER of $10^{-2}$ over} the proposed SSF-based transmission with $M=2^{10}$ and THP. Considering the data rates associated with the two schemes in Table~\ref{tab:BitRateVsConstSize}, this suggests that using a wider bandwidth is an effective means for improving the data rate. While this would be an obvious statement in general, here we note that we only make use of spectrum skirts when increasing bandwidth, which does not incur additional costs for spectrum but calls for (pre-)equalization to deal with the pulse shaping ISI.

\subsubsection{THP-DPLL for SSF-based Transmission}

We now proceed to the practical case of transmission with PN impairment. We consider the DPLL-based phase tracking and compensation as introduced in Section~\ref{NonlinearPrecoding_DPLL_CPE}. We initialize the DPLL with a training sequence and then transmit one pilot symbol in each frame of 50 symbols to aid the DPLL in maintaining its lock. 
\begin{figure}[t]
	\centering
	\subfloat[\textcolor{black}{Symbol error rate.}]{%
		\includegraphics[width=1\columnwidth]{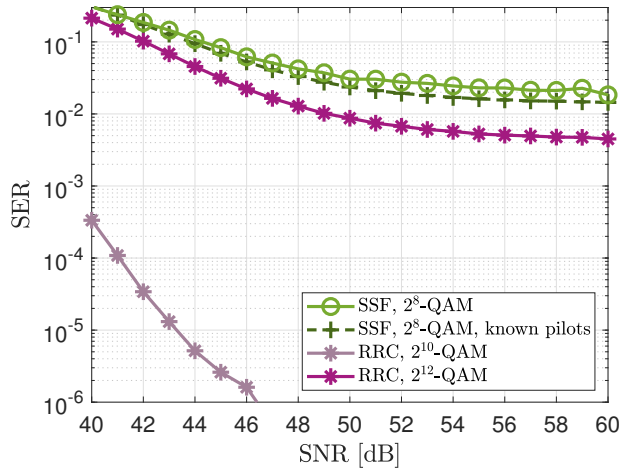}}
		\hfill
	\subfloat[\textcolor{black}{Achievable information rate.}]{%
		\includegraphics[width=1\columnwidth]{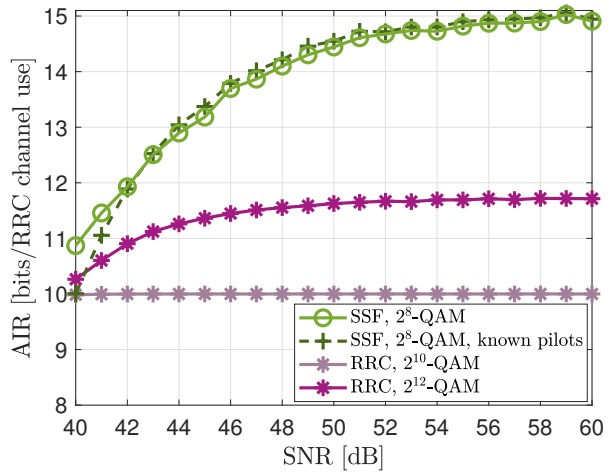}}
\caption{\textcolor{black}{SER and AIR vs.\ SNR for  SSF pulse-shaped transmission with \(2^{8}\text{-QAM}\), THP-DPLL, and PN level of  $-90$~dBc/Hz. Also included are reference curves for (i) SSF transmission and THP-DPLL assuming known effective pilot symbols in the extended constellation and (ii) conventional RRC transmission with DPLL for \(\{2^{10},\;2^{12}\}\text{-QAM}\).}}
	\label{SER_AIR_THP_DPLL_SSF_vs_RRC}
\end{figure}

Figure~\ref{SER_AIR_THP_DPLL_SSF_vs_RRC}(a) shows the SER performance of a \(2^{8}\text{-QAM}\) SSF and a \textcolor{black}{\(\{2^{10},\;2^{12}\}\text{-QAM}\)} benchmark RRC transmission, respectively. We observe an error floor of the SER curve for 
\textcolor{black}{the SSF-shaped transmission} due to PN. The pilot detection in the extended constellation as proposed in \eqref{e:ukHat} only incurs a small loss compared to the idealized case of known pilots. However, the residual phase error after DPLL-based compensation in the extended constellation before the modulo operation causes the error floor.  The conventional RRC-based transmission is also affected by PN when compensated with the DPLL. In particular, we observe the deterioration of the SER with increasing constellation size $M$. 

The translation of the SER results from Figure~\ref{SER_AIR_THP_DPLL_SSF_vs_RRC}(a) into achievable rate when error-correction coding is applied is shown in Figure~\ref{SER_AIR_THP_DPLL_SSF_vs_RRC}(b). In this figure, and in the following \textcolor{black}{ones}, AIR is presented in terms of bit per channel use of the RRC-based transmission, where the interval for one channel use is $\Trrc=1/(25.6\cdot 10^6)\approx 39$~ns. It can be seen that the proposed wideband transmission provides an improvement \textcolor{black}{of 24\%-27\%} in achievable rate for SNR values at about 50~dB and above. We note that increasing the constellation size for RRC-based transmission is not a viable option \textcolor{black}{because the transmission with \(2^{12}\text{-QAM}\) does not reach the highest number of information bits per channel use supported by the constellation order}.

\subsubsection{THP-BCJR for SSF-based Transmission}

Phase tracking with DPLL has the advantage of computational simplicity. However, we expect the performance to improve when applying the more sophisticated BCJR-based PN estimation as introduced in Section~\ref{NonlinearPrecoding_BCJR_CPE}. The PN variance is $\sigma_\psi^2 = 7.7\cdot 10^{-6}$~rad$^2$ for the symbol period $T_{\text{SSF}}\approx 20$~ns for the SSF-based transmission, and we set the maximum phase of the PN span interval as $\phimax=3.5\sqrt{\sigma_\psi^2d_{\text{pilot}}}\approx0.07$~rad. We empirically set the number of trellis nodes to 101 for a PN resolution of $1.4 \cdot 10^{-3}$~rad. 

\begin{figure}[t]
	\centering
	\subfloat[\textcolor{black}{Symbol error rate.}]{%
		\includegraphics[width=1\columnwidth]{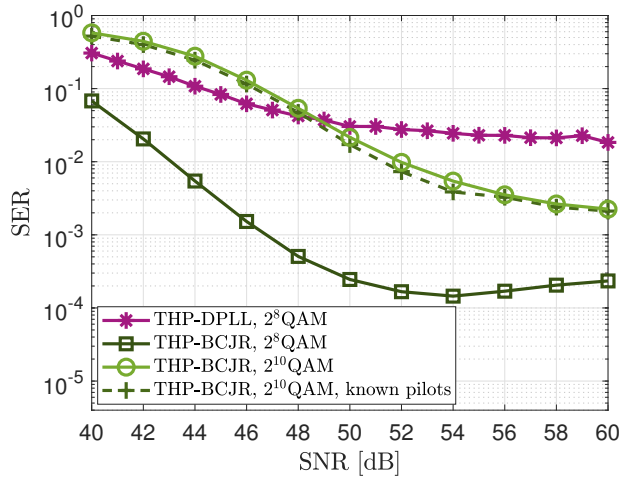}}
		\hfill
	\subfloat[\textcolor{black}{Achievable information rate.}]{%
		\includegraphics[width=1\columnwidth]{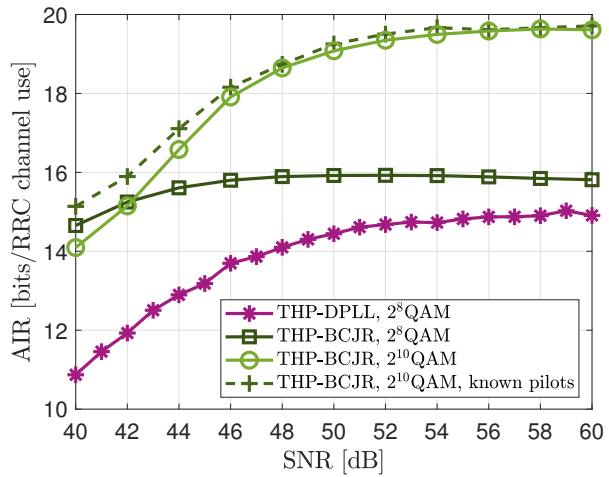}}
\caption{\textcolor{black}{SER and AIR vs.\ SNR for SSF pulse-shaped transmission for a \(\{2^{8},\;2^{10}\}\text{-QAM}\) with THP-BCJR and THP-DPLL. As a reference curve the SER result for $2^{10}$-QAM and THP-BCJR assuming known effective pilot symbols in the extended constellation is included. Phase noise level of $-90$~dBc/Hz.}}
	\label{SER_AIR_SSF_DPLL_vs_BCJR}
\end{figure}

Figure~\ref{SER_AIR_SSF_DPLL_vs_BCJR} compares the SER and AIR results for SSF-based transmission with THP-DPLL and with THP-BCJR for joint precoding and PN compensation. To obtain the PN estimates at pilot locations for THP-BCJR, we apply the decision rule in~\eqref{e:CoarsePNEstSearchSpace}. The effectiveness of this approach is inferred from the SER results for $2^{10}$-QAM  in Figure~\ref{SER_AIR_SSF_DPLL_vs_BCJR}(a), which show an almost perfect overlap of the THP-BCJR with estimation of the pilots in the extended constellation and the idealized case of known effective pilots. Next, considering the case of $2^8$-QAM transmission, the superiority of THP-BCJR over THP-DPLL can be seen in Figure~\ref{SER_AIR_SSF_DPLL_vs_BCJR}(a). In particular, the error floor experienced by THP-DPLL is lowered tremendously. Alternatively, THP-BCJR permits the use of larger signal constellations such as $2^{10}$-QAM, while still obtaining lower SERs than THP-DPLL. The improved performance in terms of uncoded SER leads to notably larger achievable rates, as shown in Figure~\ref{SER_AIR_SSF_DPLL_vs_BCJR}(b). For the $2^8$-QAM constellations, AIR improvements over a large SNR range are accomplished with THP-BCJR  compared to THP-DPLL. Furthermore, the possible use of the $2^{10}$-QAM constellation size afforded by THP-BCJR even in the presence of strong PN leads to increased maximum achievable rates. 

Next, we compare the THP-BCJR SSF-shaped transmission with the BCJR-based PN compensation applied on the benchmark RRC transmission to highlight the gain brought by utilizing the spectral skirts in terms of achievable rate. Figure~\ref{AIR_BCJR_and_overall_data_rate}(a) shows the corresponding AIR curves. The results demonstrate that increasing the achievable rate through using a higher bandwidth and shaping the signal with the SSF filter substantially outperforms the transmission with the conventional RRC filter and larger constellation sizes. In particular, increasing the QAM constellation size for the latter causes an early saturation below the maximum spectral efficiency value of $\log_2(M)$ bit/(channel use), i.e., 16 for $M=2^{16}$. This performance limiting effect of PN can to a significant extent be overcome with the proposed scheme.

\subsubsection{Overall Data-Rate Comparison}

\begin{figure}[!t]
	\centering
	\subfloat[\textcolor{black}{AIR vs.\ SNR corresponding to the SSF link with THP-BCJR for pre-equalization and PN estimation, and to the RRC link with BCJR PN estimation.}]{%
		\includegraphics[width=1\columnwidth]{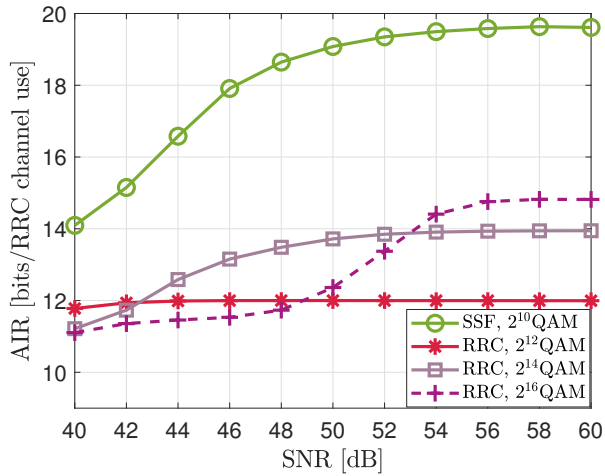}}
		\hfill
	\subfloat[\textcolor{black}{AIR (in Mbps) vs.\ SNR for the SSF and RRC transmissions where only the highest AIR for the different QAM constellations is shown for each scheme and each SNR point.}]{%
		\includegraphics[width=1\columnwidth]{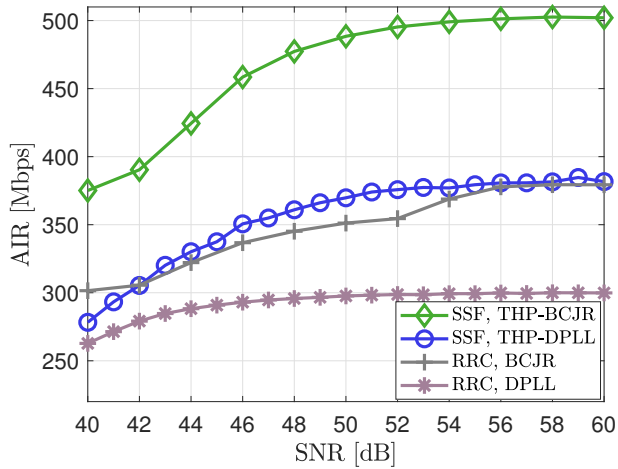}}
\caption{\textcolor{black}{AIR vs.\ SNR for (i) SSF pulse-shaped transmission with \(\{2^{8},\; 2^{10}\}\text{-QAM}\) using THP-DPLL or THP-BCJR for pre-equalization and PN estimation, and (ii) conventional RRC pulse-shaped transmission with \(\{2^{10},\;2^{12},\;2^{14},\;2^{16}\}\text{-QAM}\) using DPLL or BCJR for PN estimation. Phase noise level of $-90$~dBc/Hz.}}
	\label{AIR_BCJR_and_overall_data_rate}
\end{figure}

We provide an overall comparison of the different transmission schemes by
\textcolor{black}{illustrating envelope rate curves using} the associated AIR results in Figure~\ref{AIR_BCJR_and_overall_data_rate}(b). For easier interpretation, AIR is shown in Mbps. We vary the QAM constellation size between $2^8$ and $2^{10}$ for SSF-based transmission and between \(\{2^{10},\;2^{12},\;2^{14},\;2^{16}\}\) for RRC-based transmission, and show the highest AIR for every SNR. This mimics constellation adaptation as a function of SNR. The results in Figure~\ref{AIR_BCJR_and_overall_data_rate}(b) show the benefit of wideband transmission with pulse shaping for the entire considered SNR range. Considering the same type of PN compensation method, the proposed SSF-based transmission provides rate gains from about 10\% to \textcolor{black}{40\%} for SNR values of 40~dB and above. These are quite substantial improvements that come essentially for free as far as spectrum resources are concerned and also do not put an inordinate strain on computational complexity as only THP filtering is added compared to the conventional transmission scheme.

\subsubsection{\textcolor{black}{Coded Rate Performance}}

\begin{figure}[t]
	\centering
	\subfloat[\textcolor{black}{Achievable information rate.}]{%
		\includegraphics[width=1\columnwidth]{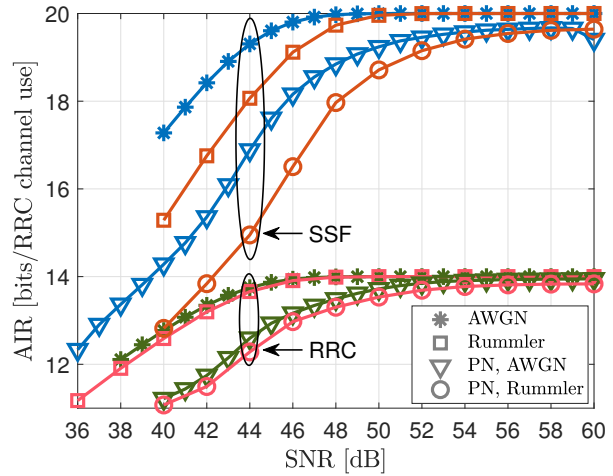}}
		\hfill
	\subfloat[\textcolor{black}{Bit error rate.}]{%
		\includegraphics[width=1\columnwidth]{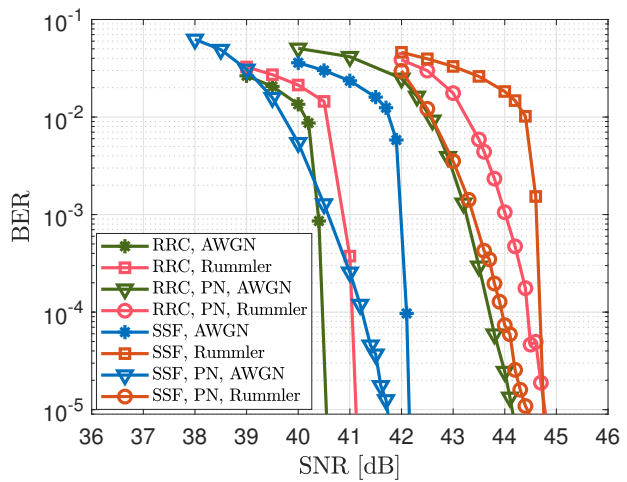}}
\caption{\textcolor{black}{AIR and BER vs.\ SNR for the SSF coded transmission with \(2^{10}\text{-QAM}\), and the conventional RRC coded transmission with \(2^{14}\text{-QAM}\). Each of the two transmissions experiences (i) AWGN channel, (ii) dispersive Rummler channel, (iii) PN, (iv) PN and dispersive Rummler channel. Rummler channel notch depth $-13$~dB, PN level of $-90$~dBc/Hz. LDPC code rate of $9/10$ for ideal carrier generation for both RRC and SSF, $5/6$ and $2/3$ for the PN impaired transmission with RRC and SSF, respectively. Symbol rates are $\Rssf=51.2$~Msymbol/s and $\Rrrc=25.6$~Msymbol/s.}}
	\label{AIR_CodedBER_SSF_RRC}
\end{figure}

\textcolor{black}{Lastly, we simulate the FEC coded BER performance for the SSF-shaped and the conventional RRC transmissions.
We apply the practical FEC coding from the DVB-S2 standard \cite{ETSIDVBS2Standard} consisting of a concatenation of a low-density parity-check (LDPC) code and a Bose–Chaudhuri–Hocquenghem (BCH) code. We consider transmissions with incrementally increasing impairments in the form of (a) one dispersive Rummler channel instance (``Rummler''), (b) PN (``PN, AWGN''), (c) PN and one dispersive Rummler channel instance (``PN, Rummler'').
We use the AIR curves in Figure~\ref{AIR_CodedBER_SSF_RRC}(a) to set the code rates such that the BER curves are expected to drop in a region around 42~dB, which allows us to show all the BER curves in one plot. In particular, the LDPC code rate is $9/10$ for RRC and SSF transmission without PN, and $5/6$ and $2/3$ for the PN-impaired transmission with RRC and SSF, respectively, and the BCH  rates are larger than 0.996. 
}

\textcolor{black}{The simulated BER curves are shown in Figure~\ref{AIR_CodedBER_SSF_RRC}(b). Firstly, we observe that the FEC coding effectively removes the high error floor exhibited by the SER in Figure~\ref{SER_AIR_SSF_DPLL_vs_BCJR}.
Secondly, since the purpose of the BER results is to validate the predictions from the AIR analysis, we consider the SNR gaps between RRC and SSF transmission at a BER of $10^{-5}$ in Figure~\ref{AIR_CodedBER_SSF_RRC}(b) and compare them with those predicted by the AIR curve in Figure~\ref{AIR_CodedBER_SSF_RRC}(a) for the corresponding data rates.\footnote{\textcolor{black}{Please note that the data rates for the scenarios in Figure~\ref{AIR_CodedBER_SSF_RRC}(b) are different as the code rates were adjusted such that BER curves drop at around 42~dB.}} Careful inspection of the numerical results shows a close match between the predicted and actual SNR gaps.
In particular, for transmission without PN the actual SNR gap in Figure~\ref{AIR_CodedBER_SSF_RRC}(b) is within 0.1~dB to 0.35~dB of the predicted one. For the PN-impaired transmissions, the actual SNR gap is within 0.85~dB to 1~dB from the predicted one, which is still fairly accurate. This confirms that inferences about the benefits of the proposed SSF transmission based on AIR results are indeed applicable when using practical FEC codes. Furthermore, the less steep drop-off of BER curves for the transmission scenarios with PN that is experienced by both the RRC and SSF transmission indicate some distortion of the FEC inputs due to PN. Hence, the simulated error rate curves suggest that a more effective PN estimation through, for example, a larger number of states in the BCJR PN estimation or processing of FEC inputs to account for mismatch (e.g.,\ \cite{nguyen:2011,alvarado:2016}) could be beneficial.}

\section{Conclusions}\label{Conclusions}
In this paper, we proposed an efficient spectrum utilization transmission suitable for microwave links. The main idea is to use spectrum skirts for wideband transmission complying with the regulated emission mask so as to avoid the need for additional frequency licenses. The main observation is that data transmission in these skirts is possible due to the frequent vacancy of adjacent channels in microwave backhaul bands. We developed a pulse shaping design and identified intersymbol interference as the primary challenge for reliable wideband transmission using spectrum skirts. We argued that pre-equalization via Tomlinson-Harashima precoding is the method of choice to deal with the known interference due to pulse shaping, which can be easily complemented by linear equalization for intersymbol interference resulting from the channel. We have combined pre-equalization with two receiver-based phase noise compensation methods, which provide different performance-complexity trade-offs. The numerical performance results in terms of uncoded symbol error rate, achievable information rate, \textcolor{black}{and coded bit error rate} demonstrated the substantial potential in terms of data rate improvements for the proposed scheme in realistic scenarios. We believe the presented transmission is an effective albeit opportunistic tool for service providers to increase their link capacities without further investments. 

\bibliographystyle{IEEEtranTCOM}
\bibliography{IEEEabrv,Bibliography_ESU}

\end{document}